# 3-Dimensional Lagrange Code for Metal (Gold) Cone Plasma

Shigeo Kawata

Utsunomiya University
Yohtoh 7-1-2, Utsunomiya 321-8585, Japan. kwt@cc.utsunomiya-u.ac.jp

**Abstract**

The document describes a numerical algorithm for plasms and fluids by the Lagrange method, in which the spatial meshes follow the plasma and fluid behavior. Through the mesh wall the plasma and the fluid do not escape. The 3D Lagrange code is originally designed to simulate a Gold cone plasma, which is relevant to confine an imploding fuel in the cone in inertial confinement fusion. However, the 3D Lagrange code would be applied to simulate plasmas and fluids, which are not deformed seriously. The 3D code is designed in the spatial Cartesian coordinate (*x, y, z*), and employs the compressible fluid model.

## 1. Introduction

The Lagrange Code is for simulations of a metal cone wall plasma, which may surround an inner DT fusion fuel shell. The metal (gold) cone protects the DT fuel or shell expansion in the transverse direction. The Lagrange code is designed to be combined with another DT-fuel implosion code, for example, O-SUKI code [1] or so.

A compressible fluid model is used to describe the cone plasma behavior. In addition to the followings, if the three-temperature model [4] or the precise radiation transport is required [5], additional equations and algorithms are needed further.

In the succeeding sections, we describe the basic equations and their discretization in the Cartesian coordinate (*x, y, z*). The following numerical algorithm is straightforward to construct the Lagrange 3D code. The Cartesian coordinate (*x, y, z*) is mapped to the logical space of $(\xi, \eta, \zeta)$. We use the mapping function between the two coordinates. Each spatial volume mesh shape is a hexahedron, which 8 vertexes. During the Lagrange mesh movement each hexahedron would be deformed. In the present numerical method, we assume all the deformed volume is described by a hexahedron volume. Therefore, the large deformation cannot be described by the following method presented in the documents.

## 2. Basic equations and discretization

The basic equations are below:

$$\frac{d\rho}{dt} = \rho \nabla \cdot \vec{v} \quad (1)$$

$$\rho \frac{d\vec{v}}{dt} = -\nabla(P + Q) \quad (2)$$

$$c_V \frac{dT}{dt} + c_T \frac{d\rho}{dt} = -\frac{1}{\rho}(P + Q)\nabla \cdot \vec{v} \quad (3)$$

Here $d/dt = \partial/\partial t + (\vec{v} \cdot \nabla)$ is the Lagrange derivative, $\rho$ the plasma mass density, $\vec{v}$ the velocity, $P$ the pressure, $Q$ the artificial viscosity to describe a shock wave in a plasma, $c_V$ the specific heat and $c_T$ the compressibility. In the Lagrange description, the equation of continuity Eq. (1) is replaced by the simple mass conservation in each computational cell:

$$M_{i,j,k} = V_{i,j,k}\rho_{i,j,k} = constant \quad (4)$$

Here $\rho_{i,j,k}$ is defined at the cell center, $V_{i,j,k}$ is the volume of each cell, $M_{i,j,k}$ the mass contained in the cell volume, and $M_{i,j,k}$ should be conserved during the computation in the Lagrangian method [2].

$$\rho_{i,j,k}{}^{n+1} = M_{i,j,k}/V_{i,j,k}{}^{n+1} \quad (4)'$$

In Fig. 1, the spatial position $\vec{x}$ in each cell is described by the following linear interpolation function from the vertex positions in the (*x, y, z*) space. The volume mesh center is indicated by a dot with "*c*".

$$\vec{x} = \xi'\eta'\zeta'\vec{x}_{i+1,j+1,k+1} + \xi'(1-\eta')(1-\zeta')\vec{x}_{i+1,j,k} + (1-\xi')\eta'\zeta'\vec{x}_{i,j+1,k+1} + \xi'\eta'(1-\zeta')\vec{x}_{i+1,j+1,k} + (1-\xi')(1-\eta')(1-\zeta')\vec{x}_{i,j,k} + (1-\xi')\eta'(1-\zeta')\vec{x}_{i,j+1,k} + (1-\xi')(1-\eta')\zeta'\vec{x}_{i,j,k+1} + \xi'(1-\eta')\zeta'\vec{x}_{i+1,j,k+1} \quad (5)$$

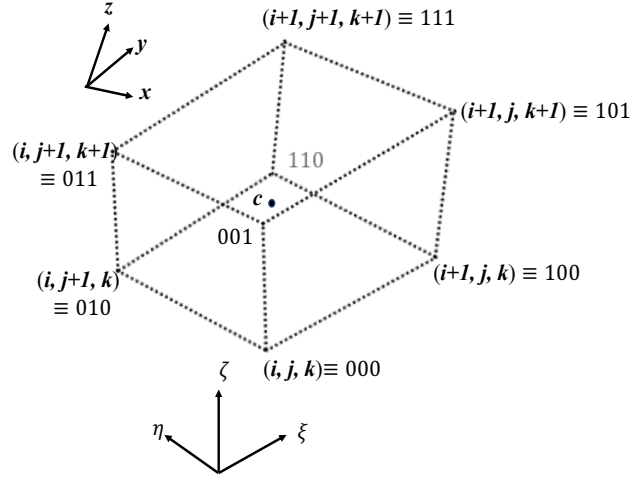

Fig. 1 An example mesh employed in the 3D Lagrange code. In the logical space each mesh is numbered by (*i, j, k*) in the logical space $(\xi, \eta, \zeta)$. The vertex (*i, j, k*) is also specified by the three digits of 000 in this paper. For example, $V_{000}$ shows $V_{i,j,k}$ and $M_{101}$ is $M_{i+1,j,k+1}$. The volume mesh center is indicated by a dot with "*c*".

Here $\xi' = \xi - i$, $\eta' = \eta - j$ and $\zeta' = \zeta - k$. The vertexes are denoted by $\vec{x}_{I,J,K}$, where $I = i$ or $i + 1$, $J = j$ or $j + 1$ and $K = k$ or $k + 1$. The position $(\xi, \eta, \zeta)$ in the logical space is inside the cell concerned and corresponds to the spatial position $\vec{x}$ in the (*x, y, z*) space. The linear interpolation function in Eq. (5) presents a volume weighting method. Equation (5) is rewritten as follows based on the shape function of $S_V(\xi', \eta', \zeta')$ at the mesh (*i, j, k*):

$$\vec{x} = \sum_{i,j,k=0}^{1} S_V(\xi', \eta', \zeta') \vec{x}_{i,j,k} \quad (6)$$

For example, $S_V(1,1,1) = \xi'\eta'\zeta'$, $S_V(1,0,0) = \xi'(1-\eta')(1-\zeta')$, $S_V(0,0,0) = (1-\xi')(1-\eta')(1-\zeta')$ and so on.

In order to calculate the volume and the mass of each mesh used in Eq. (4), we assume that each mesh size small enough, and $\rho_{i,j,k}$ is constant in each cell.

$$V_{i,j,k} = h_1 h_2 h_3 \quad (7)$$

$$h_1^2 = \left(\frac{\partial x}{\partial \xi}\right)^2 + \left(\frac{\partial y}{\partial \xi}\right)^2 + \left(\frac{\partial z}{\partial \xi}\right)^2 \quad (8)$$

$$h_2^2 = \left(\frac{\partial x}{\partial \eta}\right)^2 + \left(\frac{\partial y}{\partial \eta}\right)^2 + \left(\frac{\partial z}{\partial \eta}\right)^2 \quad (9)$$

$$h_3^2 = \left(\frac{\partial x}{\partial \zeta}\right)^2 + \left(\frac{\partial y}{\partial \zeta}\right)^2 + \left(\frac{\partial z}{\partial \zeta}\right)^2 \quad (10)$$

The volume $V_{i,j,k}$ in Eq. (7) should be evaluated at $(\xi', \eta', \zeta') = (1/2, 1/2, 1/2)$, which is the center of the mesh for the mass conservation. Each term in Eqs. (8)-(10) is driven in Appendix A.

Now we move to the equation of motion Eq. (2). The pressure $P$ is defined at the cell center "$c$". The pressure at any position $\vec{x}$ is also evaluated by

$$P(\vec{x}) = \sum_{c=1}^{8} S_c(\xi', \eta', \zeta') P_c. \quad (11)$$

Here $S_c(\xi', \eta', \zeta')$ is the similar shape function to $S_V(\xi', \eta', \zeta')$. Surrounding one vertex, where the velocity is defined, a hexahedron is formed by sub-volumes of 8 mesh volumes in Fig. 2, and the vertexes are the center points "$c$" shown in Figs. 1 and 2. One mapped figure into the $(\xi', \eta')$ space is also shown in Fig. 3.

Equation (2) is integrated over the 8 sub-volume meshes, and it assumed that the pressure $P_c$ is constant over the one volume mesh.

$$M_{V\,i,j,k} \frac{d\vec{v}}{dt} = -\int_{\Delta V} dV \, \nabla(P + Q) \quad (12)$$

Here $Q$ is the artificial viscosity to describe a shock wave, $\Delta V$ the volume of the sub-divided 8 volumes around the vertex $V$ at $(i, j, k)$ and $M_{V\,i,j,k}$ the total volume covered by $\Delta V$. In one volume mesh (see Fig. 1) the pressure $P_c$ and the artificial viscosity $Q_c$ are constant.

$$\int_{\Delta V} dV \, \nabla(P + Q) = \sum_{c=1}^{8} \nabla\{S_c(\xi', \eta', \zeta')\} dV_c \, (P_c + Q_c) \quad (13)$$

$$M_{V\,i,j,k} \frac{\vec{v}_{i,j,k}^{n+1} - \vec{v}_{i,j,k}^{n}}{dt^{n+1/2}} = -\sum_{c=1}^{8} \nabla\{S_c(\xi', \eta', \zeta')\} dV_c \, (P_c + Q_c) \quad (14)$$

Here $S_c(\xi', \eta', \zeta')$ is the shape function at the volume $\Delta V$ consisted of the sub-divided 8 volumes around the vertex $V$. The time derivative is also approximated by the time difference as shown in Fig. 4. The detail expressions in Eq. (14) are presented in the Appendix A.

The energy equation Eq. (3) is solved.

$$c_{V\,c} \frac{T_{i,j,k}^{n+1} - T_{i,j,k}^{n}}{dt^{n+1/2}} = -c_{T\,c} \frac{d\rho_c}{dt} - \frac{1}{\rho_c}(P_c + Q_c)\nabla \cdot \vec{v} \quad (15)$$

The velocity $\vec{v}(\vec{x})$ is expressed by Eq. (6) and defined at the vertex $V$:

$$\vec{v}(\vec{x}) = \sum_{i,j,k=0}^{1} S_V(\xi', \eta', \zeta') \vec{v}_{i,j,k}\big|_V = \sum_{V=1}^{8} S_V(\xi', \eta', \zeta') \vec{v}_{i,j,k}\big|_V \qquad (16)$$

Fig. 2 Example 8 meshes are displayed in the 3D space. The volume mesh center is indicated by a dot with "$c$", where the mass density $\rho_{i,j,k}$, the pressure $P$ and the temperature $P$ are defined. The plasma velocity is defined at the vertex $V$.

$$\nabla \cdot \vec{v} = \sum_{i,j,k=0}^{1} \vec{v}_{i,j,k}\big|_V \cdot \nabla S_V(\xi', \eta', \zeta') = \sum_{V=1}^{8} \vec{v}_{i,j,k}\big|_V \cdot \nabla S_V(\xi', \eta', \zeta') \qquad (17)$$

Here $\nabla S_V(\xi', \eta', \zeta') = \left(\frac{\partial S_V}{\partial x}, \frac{\partial S_V}{\partial y}, \frac{\partial S_V}{\partial z}\right)$. By Eqs. (A7)-(A9), each term is obtained. For example, Eq. (A7) for $S_V$ is below:

$$\frac{\partial S_V}{\partial x} = \frac{1}{h_1^2} \frac{\partial x}{\partial \xi} \frac{\partial S_V}{\partial \xi} + \frac{1}{h_2^2} \frac{\partial x}{\partial \eta} \frac{\partial S_V}{\partial \eta} + \frac{1}{h_3^2} \frac{\partial x}{\partial \zeta} \frac{\partial S_V}{\partial \zeta} \qquad (A7)'$$

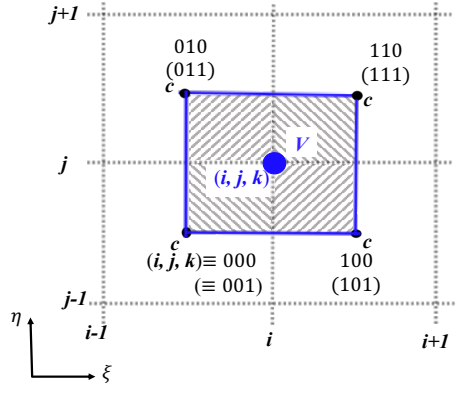

Fig. 3 2D mapping of the 3D meshes on the $\zeta' = constant$ plane. The volume mesh center is indicated by a dot with "c", where the mass density $\rho_{i,j,k}$, the pressure $P$ and the temperature $P$ are defined. The plasma velocity is defined at the vertex $V$.

Here $\left.\frac{\partial S_V}{\partial \xi}\right|_{000} = -\frac{1}{h_1^2}\frac{\partial x}{\partial \xi}(1-\eta')(1-\zeta') - \frac{1}{h_2^2}\frac{\partial x}{\partial \eta}(1-\xi')(1-\zeta') - \frac{1}{h_3^2}\frac{\partial x}{\partial \zeta}(1-\xi')(1-\eta')$. Each term of the righthand side of Eq. (A7)' is evaluated at the center of the volume mesh: $(\xi', \eta', \zeta') = (1/2, 1/2, 1/2)$. $\left.\frac{\partial S_V}{\partial \xi}\right|_{111} = \frac{1}{h_1^2}\frac{\partial x}{\partial \xi}\eta'\zeta' + \frac{1}{h_2^2}\frac{\partial x}{\partial \eta}\xi'\zeta' + \frac{1}{h_3^2}\frac{\partial x}{\partial \zeta}\xi'\eta'$.

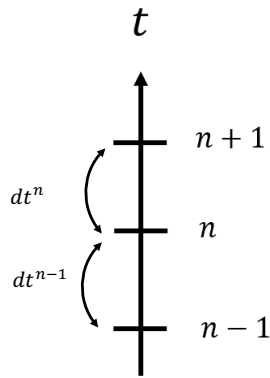

Fig. 4 Discretization of time space. The time step $dt$ is controlled to avoid the numerical instability.

$$Q_c(i,j,k) = \begin{cases} c_1{}^2 \rho_{i,j,k} \delta\vec{v}^2 + c_2{}^2 \rho_{i,j,k} c_s \, |\delta\vec{v}|, & \text{for } \delta\vec{v} < 0 \\ 0, & \text{for } \delta\vec{v} \geq 0 \end{cases} \quad (18)$$

Here $c_s$ is the sound speed, and the non-dimensional coefficients of $c_1$ and $c_2$ would be around 1.0~2.0 [3].

The artificial viscosity $Q_c$ is employed to describe shock waves in a plasma, and the kinetic energy is converted to the thermal energy.

$$\delta\vec{v}^2 = (\vec{\imath} \cdot \delta\vec{v}|_i)^2 + \left(\vec{\jmath} \cdot \delta\vec{v}|_j\right)^2 + \left(\vec{k} \cdot \delta\vec{v}|_k\right)^2 = (\delta v_i)^2 + \left(\delta v_j\right)^2 + (\delta v_k)^2 \quad (19)$$

Here $\vec{\imath} = \nabla\xi$, $\vec{\jmath} = \nabla\eta$ and $\vec{k} = \nabla\zeta$. For example, the expression $\delta\vec{v}|_i$ shows $\delta\vec{v}|_i = \sum_{j=1}^{2}\sum_{k=1}^{2}\delta\vec{v}|_{i;\,j,k} = \{(\vec{v}_{100} - \vec{v}_{000}) + (\vec{v}_{101} - \vec{v}_{001}) + (\vec{v}_{111} - \vec{v}_{011}) + (\vec{v}_{110} - \vec{v}_{010})\}/4$, which shows the velocity difference between two vertexes in the $\xi$ or $i$ direction among 8 vertexes at the volume mesh of $(i,j,k)$. The expressions for $\nabla\xi$, $\nabla\eta$ and $\nabla\zeta$ are obtained by Eqs. (8)-(10) and Eqs. (A1)-(A3). They should be evaluated at $(\xi',\eta',\zeta') = (1/2, 1/2, 1/2)$. For example, $\frac{\partial \vec{x}}{\partial \xi} = \left(\frac{\partial x}{\partial \xi}, \frac{\partial y}{\partial \xi}, \frac{\partial z}{\partial \xi}\right)$ is expressed by Eq. (A1):

$$\nabla\xi = \frac{1}{h_1^2}\left(\frac{\partial x}{\partial \xi}, \frac{\partial y}{\partial \xi}, \frac{\partial z}{\partial \xi}\right) \quad (A8)$$

$$\nabla\eta = \frac{1}{h_2^2}\left(\frac{\partial x}{\partial \eta}, \frac{\partial y}{\partial \eta}, \frac{\partial z}{\partial \eta}\right) \quad (A9)$$

$$\nabla\zeta = \frac{1}{h_3^2}\left(\frac{\partial x}{\partial \zeta}, \frac{\partial y}{\partial \zeta}, \frac{\partial z}{\partial \zeta}\right) \quad (A10)$$

$$\frac{\partial \vec{x}}{\partial \xi} = \{\eta'\zeta'\vec{x}_{111} + (1-\eta')(1-\zeta')\vec{x}_{100} + \eta'(1-\zeta')\vec{x}_{110} + (1-\eta')\zeta'\vec{x}_{101}\} - \{\eta'\zeta'\vec{x}_{011} + (1-\eta')(1-\zeta')\vec{x}_{000} + \eta'(1-\zeta')\vec{x}_{010} + (1-\eta')\zeta'\vec{x}_{001}\} \quad (A1)$$

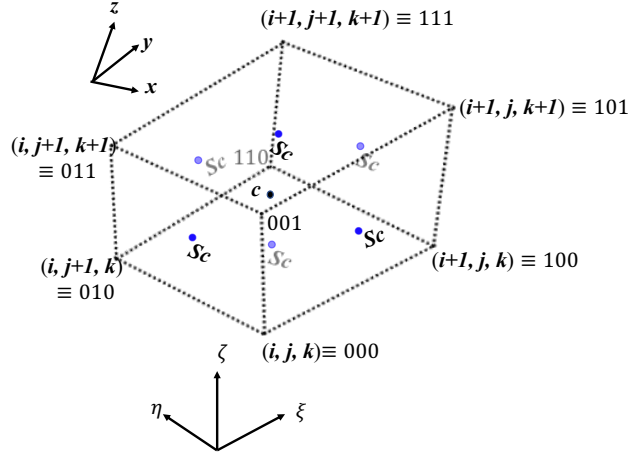

Fig. 5 An example mesh employed in the 3D Lagrange code. The heat flux is defined at the surface center indicated by "$Sc$".

If we need to include the heat conduction, Eq. (3) is modified as Eq. (20):

$$c_V \frac{dT}{dt} = \frac{1}{\rho_c} \nabla \cdot (\kappa \nabla T) \quad (20)$$

$$\nabla \cdot (\kappa \nabla T) = -\nabla \cdot \vec{F} \quad (21)$$

Here $\kappa$ is the heat conduction coefficient, and the heat flux $\vec{F}$ is $\vec{F} = -\kappa \nabla T$. The heat flux $\vec{F}$ is defined at the surface center "$Sc$" (see Fig. 5). Equation (20) is integrated over the mesh volume:

$$\iiint_{\Delta V} c_V \rho_c \frac{dT}{dt} dV \cong M_c c_V \frac{T^{n+1}-T^n}{dt} = \iiint_{\Delta V} \nabla \cdot (\kappa \nabla T) \, dV = \iint_{\Delta S} d\vec{S} \cdot \kappa \nabla T \quad (22)$$

Here the surface integral is performed on the 8 surfaces of the mesh volume concerned. If we specify the surface area vector normal to the surface by $\overrightarrow{Su_c}$, Eq. (22) is transformed as follows:

$$\iint_{\Delta S} d\vec{S} \cdot \kappa \nabla T = \sum_{Sc=1}^{8} \kappa \nabla T \cdot \overrightarrow{Su_c} = -\sum_{Sc=1}^{8} \vec{F}_{Sc} \cdot \overrightarrow{Su_c} \quad (23)$$

$$\vec{F}_{Sc} = -\kappa_{Sc} \nabla T|_{Sc} \quad (24)$$

Here $\kappa_{Sc}$ should be an averaged value of the heat conduction coefficient between two adjacent cells. For example, the temperature gradient between the mesh $(i, j, k)$ and the adjacent mesh $(i+1, j, k)$ is estimated as follows:

$$|\nabla T|_{Sc}|_{along\ i} \cong \frac{T_{i+1,j,k}-T_{i,j,k}}{\lambda_i} \qquad (25)$$

Here the distance $\lambda_i$ between the two centers of the adjacent two meshes is estimated by Eq. (6): $\lambda_i = |\vec{x}|_{c;\ i+1,j,k} - \vec{x}|_{c;\ i,j,k}|$ The direction of $\nabla T|_{Sc}$ in Eq. (25) is normal to the $\xi = constant$ surface, that is, the $i = constant$ surface. The surface area of the mesh volume $\vec{Su_c}$ is obtained as follows:

$$\vec{Su_c} = (Su_{ci}\vec{\imath}, Su_{cj}\vec{\jmath}, Su_{ck}\vec{k}) = (Su_{c\xi}\vec{\imath}, Su_{c\eta}\vec{\jmath}, Su_{c\zeta}\vec{k}) \qquad (26)$$

$$Su_{c\xi} = h_2 h_3, \quad Su_{c\eta} = h_3 h_1, \quad Su_{c\zeta} = h_1 h_2 \qquad (27)$$

Here $h_1, h_2$ and $h_3$ are shown in Eqs. (8)-(10). The vector in Eq. (26) is expressed in the space of $(\vec{\imath}, \vec{\jmath}, \vec{k})$=( $\nabla\xi$, $\nabla\eta$, $\nabla\zeta$). Each vector is again shown below:

$$\vec{\imath} = \nabla\xi = \frac{1}{h_1^2}\left(\frac{\partial x}{\partial \xi}, \frac{\partial y}{\partial \xi}, \frac{\partial z}{\partial \xi}\right) \qquad (A8)$$

$$\vec{\jmath} = \nabla\eta = \frac{1}{h_2^2}\left(\frac{\partial x}{\partial \eta}, \frac{\partial y}{\partial \eta}, \frac{\partial z}{\partial \eta}\right) \qquad (A9)$$

$$\vec{k} = \nabla\zeta = \frac{1}{h_3^2}\left(\frac{\partial x}{\partial \zeta}, \frac{\partial y}{\partial \zeta}, \frac{\partial z}{\partial \zeta}\right) \qquad (A10)$$

In order to compute the heat conduction in plasmas, usually an implicit method is recommended, because the heat conduction time scale is rather faster than the plasma movement. Usually the ADI (Alternating directional implicit) method to save the computing time and to keep the accuracy.

The equation of state [6, 7] is also required to compute the pressure $P$, the ionization degree $Z_e$, the specific heat $c_V$ and the compressibility $c_T$ from the atomic number $Z$, the atomic weight $A$, the density $\rho$ and the temperature $T$.

## 3. Algorithm review

Here we summarize the computation cycle.

1. Initial setup and computation preparations.

2. Time step $dt$ is controlled to avoid the numerical instability. $dt = C_{FL} \min\{dl/(V + C_s)\}$. The time step of $dt$ is evaluated at each mesh and each direction. The minimum $dt$ is employed. Here $V$ shows the absolute value of the plasma velocity and $C_s$ the sound speed. In plasmas and fluids shock waves may appear, and the shock speed would be larger than the sound speed. The coefficient of $C_{FL}$ should be less than 1.0. Normally we use $C_{FL} < 0.1$ to ensure the numerical stability and the numerical accuracy.

3. The new volume of each volume mesh $V_{i,j,k}{}^{n+1}$ is computed. The mass density is first obtained:

$$\rho_{i,j,k}{}^{n+1} = M_{i,j,k}/V_{i,j,k}{}^{n+1} \qquad (4)'$$

4. Then the velocity $\vec{v}_{i,j,k}^{n+1}$ is obtained at each vertex:

$$M_{V\,i,j,k} \frac{\vec{v}_{i,j,k}^{n+1} - \vec{v}_{i,j,k}^{n}}{dt^{n+1/2}} = -\sum_{c=1}^{8} \nabla\{S_c(\xi',\eta',\zeta')\}dV_c\,(P_c + Q_c) \quad (14)$$

5. The new temperature $T_{i,j,k}^{n+1}$ is obtained:

$$c_{V_c}\frac{T_{i,j,k}^{n+1} - T_{i,j,k}^{n}}{dt^{n+1/2}} = -c_{T_c}\frac{d\rho_c}{dt} - \frac{1}{\rho_c}(P_c + Q_c)\nabla\cdot\vec{v} \quad (15)$$

If needed, the thermal conduction may be computed. The energy equation may be expanded to three temperature model [4] or another precise model [5].

6. The pressure $P_c$, the specific heat $c_{V_c}$, the compressibility $c_{T_c}$ and the ionization degree are computed by the equation of state. The artificial viscosity $Q_c$ is also computed.
7. The new mesh positions are updated.

## 3. Summary

The document presents a computing algorithm for the Lagrange method to compute plasmas and fluid in 3D. Originally the algorithm is summarized for a heavy gold cone simulation. However, the algorithm can be applied to simulate plasmas and fluid generally, as far as the deformation of plasmas and fluids is limited. In the Lagrange method the volume meshes move with the plasma behavior, and the significant deformation of spatial meshes induces the mesh collapse and the computation crash.

**Acknowledgements**

The works was partly supported by JSPS, MEXT and collaborations with friends, including scientists in Utsunomiya University, Japan, Shanghai Jiao Tong University, China and ELI-Beamlines, Czech Republic.

# Appendix
## A. Derivative formulae

$$h_1^2 = \left(\frac{\partial \vec{x}}{\partial \xi}\right)^2 = \left(\frac{\partial x}{\partial \xi}\right)^2 + \left(\frac{\partial y}{\partial \xi}\right)^2 + \left(\frac{\partial z}{\partial \xi}\right)^2 \quad (8)$$

$$h_2^2 = \left(\frac{\partial \vec{x}}{\partial \eta}\right)^2 = \left(\frac{\partial x}{\partial \eta}\right)^2 + \left(\frac{\partial y}{\partial \eta}\right)^2 + \left(\frac{\partial z}{\partial \eta}\right)^2 \quad (9)$$

$$h_3^2 = \left(\frac{\partial \vec{x}}{\partial \zeta}\right)^2 = \left(\frac{\partial x}{\partial \zeta}\right)^2 + \left(\frac{\partial y}{\partial \zeta}\right)^2 + \left(\frac{\partial z}{\partial \zeta}\right)^2 \quad (10)$$

Here $\vec{x} = (x, y, z)$. From Eq. (5), each term is easily driven for the righthand sides of Eqs. (8)-(10):

$$\frac{\partial \vec{x}}{\partial \xi} = \{\eta'\zeta'\vec{x}_{111} + (1-\eta')(1-\zeta')\vec{x}_{100} + \eta'(1-\zeta')\vec{x}_{110} + (1-\eta')\zeta'\vec{x}_{101}\} -$$
$$\{\eta'\zeta'\vec{x}_{011} + (1-\eta')(1-\zeta')\vec{x}_{000} + \eta'(1-\zeta')\vec{x}_{010} + (1-\eta')\zeta'\vec{x}_{001}\} \quad (A1)$$

$$\frac{\partial \vec{x}}{\partial \eta} = \{\xi'\zeta'\vec{x}_{111} + (1-\xi')\zeta'\vec{x}_{011} + \xi'(1-\zeta')\vec{x}_{110} + (1-\xi')(1-\zeta')\vec{x}_{010}\} -$$
$$\{\xi'(1-\zeta')\vec{x}_{100} + (1-\xi')(1-\zeta')\vec{x}_{000} + (1-\xi')\zeta'\vec{x}_{001} + \xi'\zeta'\vec{x}_{101}\} \quad (A2)$$

$$\frac{\partial \vec{x}}{\partial \zeta} = \{\xi'\eta'\vec{x}_{111} + (1-\xi')\eta'\vec{x}_{011} + (1-\xi')(1-\eta')\vec{x}_{001} + \xi'(1-\eta')\vec{x}_{101}\} -$$
$$\{\xi'(1-\eta')\vec{x}_{100} + \xi'\eta'\vec{x}_{110} + (1-\xi')(1-\eta')\vec{x}_{000} + (1-\xi')\eta'\vec{x}_{010}\} \quad (A3)$$

In Eq. (11) of $P(\vec{x}) = \sum_{c=1}^{8} S_c(\xi', \eta', \zeta') P_c$, $S_c(\xi', \eta', \zeta')$ is easily derived as follows. By Eq. (6) the center point coordinates are derived.

$$\vec{x}|_c = \sum_{i,j,k=0}^{1} S_V\left(\xi' = \frac{1}{2}, \eta' = \frac{1}{2}, \zeta' = \frac{1}{2}\right) \vec{x}_{i,j,k} \quad (6)$$

At each volume mesh the center point corresponds to the logical point of $(\xi', \eta', \zeta') = (1/2, 1/2, 1/2)$. First, each central point is obtained for all 8 meshes surrounding the vertex $V$ of $(i, j, k)$. Then the sub-volume $dV_c$ is calculated.

$$dV_c|_{ijk} \cong (h_1 h_2 h_3)|_{ijk}/8 \quad (A4)$$

$$dV_c = \sum_{c=1}^{8} dV_c|_{ijk} = \sum_{i=i-1}^{i} \sum_{j=j-1}^{j} \sum_{k=k-1}^{k} dV_c|_{ijk} \quad (A5)$$

In the equation of motion Eq. (13) requires $\nabla S_c(\xi', \eta', \zeta') = \left(\frac{\partial S_c}{\partial x}, \frac{\partial S_c}{\partial y}, \frac{\partial S_c}{\partial z}\right)$.

$$P(\vec{x}) = \sum_{c=1}^{8} S_c(\xi', \eta', \zeta') P_c$$

$$\cong \xi'\eta'\zeta' P_{c_{i+1,j+1,k+1}} + \xi'(1-\eta')(1-\zeta') P_{c_{i+1,j,k}}$$

$$+ (1-\xi')\eta'\zeta' P_{c_{i,j+1,k+1}} + \xi'\eta'(1-\zeta') P_{c_{i+1,j+1,k}}$$

$$+ (1-\xi')(1-\eta')(1-\zeta') P_{c_{i,j,k}} + (1-\xi')\eta'(1-\zeta') P_{c_{i,j+1,k}}$$

$$+ (1-\xi')(1-\eta')\zeta' P_{c_{i,j,k+1}} + \xi'(1-\eta')\zeta' P_{c_{i+1,j,k+1}}$$

$$= \xi'\eta'\zeta' P_{c_{111}} + \xi'(1-\eta')(1-\zeta') P_{c_{100}} + (1-\xi')\eta'\zeta' P_{c_{011}} + \xi'\eta'(1-\zeta') P_{c_{110}} + (1-\xi')(1-\eta')(1-\zeta') P_{c_{000}} + (1-\xi')\eta'(1-\zeta') P_{c_{010}} + (1-\xi')(1-\eta')\zeta' P_{c_{001}} + \xi'(1-\eta')\zeta' P_{c_{010}}$$

(A6)

Based on Eq. (A6), $\nabla S_c(\xi',\eta',\zeta') = \left(\frac{\partial S_c}{\partial x}, \frac{\partial S_c}{\partial y}, \frac{\partial S_c}{\partial z}\right)$ is derived:

$$\frac{\partial S_c}{\partial x} = \frac{1}{h_1^2}\frac{\partial x}{\partial \xi}\frac{\partial S_c}{\partial \xi} + \frac{1}{h_2^2}\frac{\partial x}{\partial \eta}\frac{\partial S_c}{\partial \eta} + \frac{1}{h_3^2}\frac{\partial x}{\partial \zeta}\frac{\partial S_c}{\partial \zeta} \quad (A7)$$

$$\frac{\partial S_c}{\partial y} = \frac{1}{h_1^2}\frac{\partial y}{\partial \xi}\frac{\partial S_c}{\partial \xi} + \frac{1}{h_2^2}\frac{\partial y}{\partial \eta}\frac{\partial S_c}{\partial \eta} + \frac{1}{h_3^2}\frac{\partial y}{\partial \zeta}\frac{\partial S_c}{\partial \zeta} \quad (A8)$$

$$\frac{\partial S_c}{\partial z} = \frac{1}{h_1^2}\frac{\partial z}{\partial \xi}\frac{\partial S_c}{\partial \xi} + \frac{1}{h_2^2}\frac{\partial z}{\partial \eta}\frac{\partial S_c}{\partial \eta} + \frac{1}{h_3^2}\frac{\partial z}{\partial \zeta}\frac{\partial S_c}{\partial \zeta} \quad (A9)$$

For example, $\nabla S_c(\xi',\eta',\zeta')|_{i,j,k=000} = \left(\frac{\partial S_c}{\partial x}, \frac{\partial S_c}{\partial y}, \frac{\partial S_c}{\partial z}\right)\Big|_{000}$ is shown below. $S_c(\xi',\eta',\zeta')|_{000} = (1-\xi')(1-\eta')(1-\zeta')$. $\frac{\partial S_c}{\partial x}\Big|_{000} = -\frac{1}{h_1^2}\frac{\partial x}{\partial \xi}(1-\eta')(1-\zeta') - \frac{1}{h_2^2}\frac{\partial x}{\partial \eta}(1-\xi')(1-\zeta') - \frac{1}{h_3^2}\frac{\partial x}{\partial \zeta}(1-\xi')(1-\eta')$. Here $\frac{\partial S_c}{\partial x}\Big|_{000} = \left(-\frac{1}{4h_1^2}\frac{\partial x}{\partial \xi} - \frac{1}{4h_2^2}\frac{\partial x}{\partial \eta} - \frac{1}{4h_3^2}\frac{\partial x}{\partial \zeta}\right)\Big|_{000}$ is evaluated at $(\xi',\eta',\zeta') = (1/2, 1/2, 1/2)$ and the right hand side is obtained from Eqs. (8)-(10) and Eqs. (A1)-(A3). Another example at $(i,j,k) = (i+1, j+1, k+1) = 111$ is also shown here: $\nabla S_c(\xi',\eta',\zeta')|_{i,j,k=111} = \left(\frac{\partial S_c}{\partial x}, \frac{\partial S_c}{\partial y}, \frac{\partial S_c}{\partial z}\right)\Big|_{111}$ and $S_c(\xi',\eta',\zeta')|_{111} = \xi'\eta'\zeta'$. $\frac{\partial S_c}{\partial x}\Big|_{111} = \frac{1}{4h_1^2}\frac{\partial x}{\partial \xi} + \frac{1}{4h_2^2}\frac{\partial x}{\partial \eta} + \frac{1}{4h_3^2}\frac{\partial x}{\partial \zeta}$.

Here it should be noticed that $h_1$ is the distance between the center position "$c$" and the vertex "$V$", and for one volume mesh in Eqs. (7)-(10) $h_1$ is the distance between two vertexes. Therefore, $h_1$ in this paragraph is a half of the distance in Eqs. (7)-

(10).

Here $\nabla\xi, \nabla\eta$ and $\nabla\zeta$ show the orthogonal vectors to the $\xi = constant$ surface, the $\eta - constant$ direction and the $\zeta - constant$ direction, respectively. They are also listed below:

$$\nabla\xi = \frac{1}{h_1^2}\left(\frac{\partial x}{\partial \xi}, \frac{\partial y}{\partial \xi}, \frac{\partial z}{\partial \xi}\right) \quad (A8)$$

$$\nabla\eta = \frac{1}{h_2^2}\left(\frac{\partial x}{\partial \eta}, \frac{\partial y}{\partial \eta}, \frac{\partial z}{\partial \eta}\right) \quad (A9)$$

$$\nabla\zeta = \frac{1}{h_3^2}\left(\frac{\partial x}{\partial \zeta}, \frac{\partial y}{\partial \zeta}, \frac{\partial z}{\partial \zeta}\right) \quad (A10)$$